\font\twelvebf=cmbx12
\font\ninerm=cmr9
\nopagenumbers
\magnification =\magstep 1
\overfullrule=0pt
\baselineskip=18pt
\line{\hfil }
\line{\hfil May 1997}
\vskip .8in
\centerline{\twelvebf On The Planar Yang-Mills Theory}
\vskip .5in
\centerline{\ninerm D.MINIC}
\centerline{Physics Department}
\centerline{Pennsylvania State University}
\centerline{University Park, PA 16802}
\centerline{and}
\centerline{Enrico Fermi Institute}
\centerline{University of Chicago}
\centerline{Chicago, IL 60637}
\centerline {dminic@yukawa.uchicago.edu}

\vskip 1in
\baselineskip=16pt
\centerline{\bf Abstract}
\vskip .1in

The planar Yang-Mills theory in three spatial dimensions
is examined in a particular representation which explicitly 
embodies factorization.
The effective planar Yang-Mills theory Hamiltonian is constructed
in this representation.

\vfill\eject

\footline={\hss\tenrm\folio\hss}

\magnification =\magstep 1
\overfullrule=0pt
\baselineskip=22pt
\pageno=2
\vskip .2in

The planar Yang-Mills theory [1] (or the $N=\infty$ limit of the 
$SU(N)$ Yang-Mills theory) is characterized by a remarkable property of 
factorization, 
that is, 
${\langle}X Y{\rangle}={\langle}X{\rangle}{\langle}Y{\rangle}$, 
where $X$ and $Y$ are some gauge-invariant observables.
This statement implies "classicality" of the planar Yang-Mills theory. 
In this note I would like to propose a particular heuristic 
representation of the planar Yang-Mills theory in three spatial dimensions
which explicitly embodies factorization.
(The heuristic construction presented below is in some sense an 
extrapolation
of the well-known results of [2].) I want to 
emphasize that most of the formulae stated below should be 
understood, for the time being, as mere formal expressions.
The very important question of renormalization (after suitable
regularization) is not considered at all.
In what follows I adopt the Hamiltonian framework 
in three spatial dimensions. (The Planck 
constant is set to one.)

I would like to address the following
two questions: "What is a natural representation
of the $SU(\infty)$ gauge connections $A_{i}(\vec{x})$ such that
 the gauge structure of the $SU(\infty)$
Yang-Mills theory is fully preserved?", and "Given such a representation 
of 
$A_{i}(\vec{x})$, how does the ground state of the planar Yang-Mills 
theory
look like, and furthermore, what is the form of 
the effective planar Yang-Mills Hamiltonian?". 

In the $N=\infty$ limit the gauge connections 
$A_{i}(\vec{x})$ are $\infty$ $\times$ 
$\infty$ matrices at each space point. 
In order to make sense of this basic fact
it is absolutely essential to come up with a 
suitable representation for $A_{i}(\vec{x})$ so that the matrix
structure looks transparent. Moreover, the representation should be such 
so that the gauge structure of the Yang-Mills theory appears natural,
and that, for example, the
known perturbative results can be readily recovered following the
standard procedure. (In the one-matrix model of [2], the gauge 
transformations
are essentially similarity transformations; therefore any $\infty$ 
$\times$
$\infty$ matrix can be naturally represented by its eigenvalues. The 
representation proposed below similarly follows the general form of
the $SU(\infty)$ Yang-Mills gauge transformations).

In order to capture the basic
matrix properties of the gauge connections $A_{i}(\vec{x})$, 
as well as the corresponding
 non-abelian features of the planar theory, I start from 
the following correspondence relation
$$
A_{i}(\vec{x}) \rightarrow a_{i}(\vec{x},\alpha,\beta) ,    \eqno(1)
$$ 
where $\alpha$ and $\beta$ are
two real parameters, and $a_{i}$'s are functions of $x$, $\alpha$, 
$\beta$.
(In other words, the matrix indices, which play the role of
 internal parameters
and which in the planar limit run from zero to
infinity, are replaced by two continuous, external parameters 
$\alpha$ and $\beta$.) 
Given (1) how does one go about representing the non-abelian features 
of the theory? In particular, what is the form of a suitable 
representation of
the commutator bracket?
I propose the following identification
$$
[A_{i}(\vec{x}), A_{j}(\vec{x})] \rightarrow 
\{a_{i}(\vec{x},\alpha,\beta), a_{j}(\vec{x},\alpha,\beta)\},   \eqno(2)
$$
where $\{,\}$ denotes the Poisson bracket with respect to 
 $\alpha$ and $\beta$, or
$$
\{a_{i}(\vec{x},\alpha,\beta), a_{j}(\vec{x},\alpha,\beta)\} =
(\partial_{\alpha}a_{i}(\vec{x},\alpha,\beta)  
\partial_{\beta}a_{j}(\vec{x},\alpha,\beta) -
\partial_{\beta}a_{i}(\vec{x},\alpha,\beta)
\partial_{\alpha}a_{j}(\vec{x},\alpha,\beta)) .           \eqno(3)
$$
Thus all expressions containing the gauge connection
and the commutator are to be replaced with the "identically"
looking ones, after the translation defined
in (1) and (2) has been applied. Also, the operation of
tracing should be replaced by the operation of integration over the extra 
continuous parameters $\alpha$ and $\beta$
$$
Tr  \rightarrow   \int d\alpha d\beta.   \eqno(4)
$$

(The identical dictionary was suggested in a related
context by the authors of [3], who made use of the equivalence
between the $SU(\infty)$ Lie algebra and the algebra of
area preserving diffeomorphisms of a two-dimensional sphere $S^{2}$,
parametrized by $\alpha$ and $\beta$. This equivalence is very
important for the discussion of the ground state in what follows.)

The authors of [3] have also suggested
the following translation which leads to a natural representation
of the $SU({\infty})$ structure constants, namely
$$
A_{i}^{c}(\vec{x})t^{c}  \rightarrow 
\sum_{lm} a_{i}^{lm}(\vec{x}) 
Y_{lm}(\alpha,\beta) ,             \eqno(5)
$$
where $t^{c}$ are the generators of $SU(\infty)$ and 
$Y_{lm}(\alpha,\beta)$
are the $S^{2}$ spherical harmonics. The $SU(\infty)$ structure constants
are then identified with the structure constants of the area preserving 
diffeomorphisms of a two-sphere, defined in terms of the spherical 
harmonics basis [3]
$$
\{Y_{lm},Y_{l'm'}\} = f_{lm,l'm'}^{l''m''}Y_{l''m''}.     \eqno(6)
$$

Then the expression defining the gauge transformations  
reads as follows
$$
\delta a_{i}(\vec{x}, \alpha, \beta) = 
\partial_{i} \Omega(\vec{x}, \alpha, \beta)
+ g\{a_{i},\Omega\},                \eqno(7)
$$
and the corresponding formula for the field strength is
$$
F_{ij} = \partial_{i} a_{j} - \partial_{j} a_{i} + g\{a_{i},a_{j}\}.
\eqno(8)
$$

Given the above dictionary one can recover the standard results
of pertubation theory, such as asymptotic freedom,
starting from the Hamiltonian
$$
H = \int d{\vec{x}} d\alpha d\beta {1 \over 2}(p_{i}^{2} + F_{ij}^{2}(a)),
\eqno(9)
$$
where $p_{i} \rightarrow -i{\delta \over {\delta a_{i}}}$ 
and the magnetic field $F_{ij}$ is given by
(8). (The well-known perturbative 
results can be easily recovered in the background-field approach.)

  What is the nature of the ground state of the planar Yang-Mills theory
in view of the correspondence relations (1) - (9)? 
(The ground state being the only surviving
state according to factorization.)
In order to answer that question I wish to 
use the Hamiltonian version of the
"constrained classical dynamics" formalism [4] (applicable both to
vector and matrix planar field theories) which quite naturally
incorporates the fundamental property of factorization through
the following commutation relations between the planar gauge field
and its conjugate momentum 
$$
[A_{i}(\vec{x}),P_{j}(\vec{y})] = i \delta_{ij} \delta(\vec{x} - \vec{y})
  |0{\rangle}{\langle}0|.
\eqno(10)
$$
(In other words, in the expansion of unity that appears in the
usual canonical commutation relations 
$[A_{i}(\vec{x}),P_{j}(\vec{y})] = i \delta_{ij} \delta(\vec{x} - 
\vec{y})$
$$
1 = |0{\rangle}{\langle}0| + \sum_{n=1} |n{\rangle}{\langle}n|,
\eqno(11)
$$
only the first term, which is a projection operator, is kept. Note that
(10) can be understood as 
${\langle}0|[A_{i}(\vec{x}),P_{j}(\vec{y})]|0{\rangle}
 = i \delta_{ij} \delta(\vec{x} - \vec{y})$, which in view of (1)
implies $P_{i}(\vec{x}) \rightarrow p_{i}(\vec{x},\alpha,\beta)$.)
The dynamical equations of motion are given by the familiar expressions 
[4]
$$
i [H_{r},A_{i}(\vec{x})]=  \dot A_{i}(\vec{x}),
\eqno(12)
$$
and
$$
i [H_{r},P_{i}(\vec{x})]=  \dot P_{i}(\vec{x}).
\eqno(13)
$$
It is important to note  
that $H_{r}$ represents the reduced Hamiltonian (reduced onto the
ground state of the theory). 
Equations (10), (12) and (13) define the effective Hamiltonian version 
of the planar Yang-Mills theory.

Now I wish to consider the following concrete realization of such  
generalized quantum Hamiltonian dynamics 
based on a very particular representation of the projection operator in 
the
planar commutation relations (10):
$$
|0{\rangle}{\langle}0| = \psi^{\dagger} \psi,
\eqno(14)
$$
where $\psi^{2}={\psi^{\dagger}}^{2} = 0$, 
$\psi \psi^{\dagger} + \psi^{\dagger} \psi =1$, i.e. $\psi$ and
$\psi^{\dagger}$ are fermionic operators.
This representation is suggestive of a fermionic ground state.

That fact can be seen from the
planar commutation relations (10), given the fermionic
realization of the projector (14).
Note, that due to the fact that $\psi^{\dagger} \psi$ is a fermion
number operator, each phase cell $\Delta a_{i} \Delta p_{i}$ contains
a single fermion. (The same fermion number operator serves as a generator 
of
the area preserving diffeomorphisms of $S^{2}$, which is compatible with
(1) and (2), so the ground state
satisfies the Gauss law, that is, it is  
invariant under (7).)
Given that, I conclude that the ground state is basically
 characterized by a certain region of the functional
phase space $Da_{i} Dp_{i}$ 
which is characterized by the fundamental property
of incompressibility according to the Liouville theorem. (These 
$a_{i}$ and $p_{i}$ configurations
saturate the planar limit.)
In other words,  the following constraint
(which is compatible with the Gauss law) is imposed 
on the functional phase space volume
$$
\int Da_{i} Dp_{i} \theta(e - H) =1,   \eqno(15)
$$
where $H$ denotes the Hamiltonian (9), $e$ stands for the Fermi
energy and $\theta$ is
the usual step function. 
 
Equation (15)
tells us that the volume of the functional phase space fluid is to be 
normalized to one in such a way, as if there existed a single 
fermion placed at each phace space
cell, and consequently, taking into account 
the Pauli exclusion principle, as if there existed, in the limit
of a large number of cells,  
an incompressible fermionic fluid, with
the Fermi energy $e$.
By recalling that each phace space cell has 
a natural volume of the order of the Planck constant and that 
the planar
limit corresponds to a situation where the number of cells
goes to infinity, the product of the Planck
constant and the number of cells can be adjusted to one (the reason being 
that the $1/N$ expansion formally 
corresponds to a "semiclassical" expansion, $1/N$ acting as an effective
"Planck constant"). Hence follows the
relation (15), describing an incompressible drop
of functional phase space of unit volume. (The appearance of fermions 
could be intuitively understood from the point of view of 't Hooft's
 double-line
representation for the planar graphs [1]. The fact that such
 lines do not
cross in the planar limit is achieved by attaching fermions to each line 
and
using the exclusion principle.)  

>From this vantage point relation (15) gives a rather natural, even though
implicit realization of
the ground state of the planar Yang-Mills theory, that is compatible with 
the
Gauss law.

By formally integrating over $p_{i}(\vec{x},\alpha,\beta)$ in (15), 
 a constraint imposed on the part of the 
configuration space variables $a_{i}(\vec{x},\alpha,\beta)$
relevant for the planar limit, is obtained
$$
\int Da_{i} \rho(a) = 1,       \eqno(16)
$$
where the functional $\rho(a)$ is, again formally, defined by 
$$
\rho(a) = \int Dp_{i} \theta(e - H).   \eqno(17)
$$
In other words the functional $\rho(a)$ corresponds to the volume
of the functional momentum space that is  
relevant for the description of the ground state.
Equation (16) provides  
another suitable representation of
the ground state of the planar Yang-Mills theory, again compatible with
the Gauss law.
One could interpret (16) as
$$ 
\int Da_{i} \delta (a_{i} - A_{i}) = 1, \eqno(18)
$$
which contains the same information as the starting equation (1).

Now it might seem reasonable to introduce the weight functional $\rho(a)$
as the appropriate new variable for an effective Hamiltonian description 
of
the planar Yang-Mills theory. 
 In order to accomplish this one might adopt, for example,
 the well-known 
collective-field theory (or what is more appropriate in this
case - collective-functional theory) approach of Jevicki and Sakita [5]. 

Here I would like to use the already established fermionic-fluid picture 
of 
the ground state (15).
Then the effective planar Hamiltonian is given by
$$
H_{r} = \int Da_{i} \int Dp_{i}
( \int d^{3}x d\alpha d\beta {1 \over 2}(p_{i}^{2}  +  
{F_{ij}(a)}^{2})) \theta(e - \int d^{3}x d\alpha d\beta 
{1 \over 2}(p_{i}^{2}  +  {F_{ij}(a)}^{2})),     \eqno(19)
$$
or in terms of a fermionic functional $\Psi(a)$ which describes the
fermionic nature of the vacuum  
$$
H_{r} = \int d^{3}x d\alpha d\beta \int Da_{i} 
({1 \over 2}{\delta {\Psi^{\dagger}} \over {\delta a_{i}}}
{\delta {\Psi} \over {\delta a_{i}}} +({1 \over 2}F_{ij}^{2}(a) - e)
\Psi^{\dagger}(a) \Psi(a) ),      \eqno(20)
$$
where $\Psi^{\dagger}(a) \Psi(a)) =\rho(a)$, $\rho(a)$ being defined by
(17). This formula can be understood as the usual expression for the
ground state energy, written in a
second quantized manner, after taking into account the fact that the 
ground
state of the planar theory is fermionic, as implied by (15). 
( Given $H_{r}$ it seems very difficult to recover
 any perturbative results 
precisely because now the true, non-perturbative
 vacuum is known, being described by (15).)

Note that the above expressions for the effective 
Hamiltonian contain functional integrals, the
fact which tells us that we are not dealing with an
ordinary
field theory. (Here lies the fundamental difference between
$N=\infty$ vector and matrix field-theory models: 
The planar limit of vector field-theory
models
is described in terms of suitable (bilocal) field variables, which 
implement the summation over all "bubble" diagrams, while the planar 
limit of matrix field-theory models, such as Yang-Mills theory, is
described in terms of suitable functional variables, such as $\rho(a)$,
or $\Psi(a)$,
which implement the summation over all planar diagrams. 
In fact, the above formulation of the planar Yang-Mills theory
has a certain flavor of
what one might call "string field theory" or "non-local functional-field
theory".)

One could use the weight functional $\rho(a)$ as a collective functional
in the spirit of [5]. Then the expression (19) could be interpreted
as the effective potential of the Jevicki-Sakita collective-functional
 Hamiltonian,
after the application of the definition (17). The minimum of the
effective potential, which determines the ground state of the planar
Yang-Mills theory, is in turn given by (15). (The Fermi energy $e$
plays the role of a Lagrange multiplier, imposing the constraint
(16) in this approach.) Unfortunately, unlike in the one-matrix
model case [2], a simple explicit expression for the 
collective functional $\rho(a)$ cannot be readily obtained,
 precisely because of the functional integral over the first term 
$p_{i}^{2}$ in the expression (19) for the effective collective
functional Hamiltonian. The same term governs the physics of
elementary excitations around the ground state (15). In other words,
when expanded to second order in $\rho(a)$
around the minimum defined by (15), the same term determines the 
frequencies
of small oscillations around the minimum of the effective potential
(this meshes nicely with the intuitive picture developed in [6]).
The fundamental frequency $\omega_{0}$ is controlled by the
density of states 
$$
\sigma(e) = \int Da_{i} Dp_{i} \delta (e - H),   \eqno(21)
$$
as follows
$$
\omega_{0}^{-1} = \sigma (e).   \eqno(22)
$$
The fundamental frequency $\omega_{0}$ is, at least formally,
 positive definite. (Again, an 
explicit functional expression in terms of $a_{i}(\vec{x}, \alpha, 
\beta)$ 
is not readily available).

In conclusion, let me summarize the essential points of the above 
realization
of the planar Yang-Mills theory:

i) A suitable representation 
of the $SU(\infty)$ gauge connection is provided through (1) (along with
 the related
prescriptions for the commutators, traces and $SU(\infty)$
structure constants; the respective equations (2), (4) and (6)). This 
representation utilizes the equivalence between the $SU(\infty)$ Lie 
algebra
and the algebra of area preserving diffeomorphisms of a two-sphere [3].

ii) A suitable representation of the ground state 
of the planar Yang-Mills theory,
 that is compatible with the Gauss law,  
is provided through 
(15). This representation is based on (i) and leads to a 
fermionic-fluid picture
of the ground state.

iii) Given the fermionic-fluid picture of the ground state of the planar
Yang-Mills theory, the effective planar Yang-Mills
Hamiltonian is obtained; equations (19) and (20).

\vskip.1in
I thank V. P. Nair for crucial discussions and a critical reading of the
original version of this article. I am grateful to Shyamoli Chaudhuri,
Joseph Polchinski and
Branko Uro\v{s}evi\'{c} for comments. I also thank Bunji Sakita for 
drawing
my attention to his work with Kavalov.
I also wish to thank members of the University of Chicago Theory Group
 for their kind hospitality.
This work was fully and generously supported by 
the Joy K. Rosenthal Foundation.

\vskip.1in
{\bf References}
\item{1.} G. 't Hooft, Nucl. Phys. B72 (1974) 461; E. Witten, 
Nucl. Phys. B160 (1979) 519; for a collections of papers on large N
methods consult {\it The large N Expansion in Quantum Field Theory 
and Statistical Mechanics}, eds. E. Brezin and S. Wadia, World Scientific,
1994. 
\item{2.} E. Brezin, C. Itzykson, G. Parisi and J. B. Zuber, 
Comm. Math. Phys. 59 (1978) 35.
\item{3.} E. G. Floratos, J. Iliopoulos and G. Tiktopoulos, 
Phys. Lett. B217 (1989) 285; J. Hoppe, Ph. D. thesis (MIT, 1988);
A. Kavalov and B. Sakita, hep-th/9603024.
\item{4.} K. Bardakci, Nucl. Phys. B178 (1980) 263; 
O. Haan, Z. Phys C6 (1980) 345; M. B. Halpern, Nucl. Phys. B188 (1981) 
61;  
M. B. Halpern and C. Schwarz, Phys. Rev. D24 (1981) 2146; 
A. Jevicki and N. Papanicolaou, Nucl. Phys. B171 (1980) 363;
A. Jevicki and H. Levine, Phys. Rev. Lett 44 (1980) 1443;
 Ann. Phys. 136 (1981) 113. 
\item{5.} B. Sakita, Phys. Rev. D21, (1980) 1067; 
A. Jevicki and B. Sakita, Nucl. Phys. B165 (1980) 511; 
Nucl. Phys. B185 (1981) 89.
\item{6.} R. P. Feynman, Nucl. Phys. B188 (1981) 479.

\end